\documentclass[amsmath,amssymb]{JHEP3}
\usepackage{graphicx}
\usepackage{eufrak}
\usepackage{mathrsfs}
\usepackage{subfigure}
\title{The fluid/gravity correspondence: \\ {\small Lectures notes from the 2008 Summer School on Particles, Fields, and Strings, UBC, Canada}}
\author{Nicola Ambrosetti\\ University of Bern, ambrose@itp.unibe.ch}
\author{James Charbonneau\\ University of British Columbia, james@phas.ubc.ca}
\author{Silke Weinfurtner\\ University of British Columbia, silke.weinfurtner@ubc.ca}
\abstract{This is a paper compiled by students of the 2008 Summer School on Particles, Fields, and Strings held at the University of British Columbia on lectures given by Veronika Hubeny as understood and interpreted by the authors.  We start with an introduction to the AdS/CFT duality. More specifically, we discuss the correspondence between relativistic, conformal hydrodynamics and Einstein's theory of gravity. Within our framework the Einstein equations are an effective description for the string theory in the bulk of AdS$_5$ spacetime and the hydrodynamic fluid equations represent the conformal field theory near thermal equilibrium on the boundary. In particular we present a new technique for calculating properties in fluid dynamics using the stress-energy tensor induced on the boundary, by the gravitational field in the bulk, and comparing it with the form of the stress-energy tensor from hydrodynamics. A detailed treatment can be found in [JHEP 02 (2008) 045] and
[arXiv:0803.2526].}
\preprint{}
\begin{document}

%
\section{Introduction\label{Sec:Introduction}}
%
The motivation for studying the fluid/gravity correspondence is to gain insight into string theory and strongly coupled gauge theories.  There is a symbiotic relationship between these theories and the insight gained is rooted in the fact that they are dual to each other via the AdS/CFT correspondence. Their respective coupling constants can be matched to scale inversely to each other, meaning that doing perturbative calculations in the weak coupling limit of one theory can give us non-perturbative results in the strong coupling limit of the other.  This is provided we have the dictionary between the two theories that give us a map between objects in both theories.

One of the fundamental questions to be answered in quantum gravity is what is the fundamental nature of spacetime?  Somehow spacetime is an emergent property, but what does it emerge from?  Unfortunately, because the dictionary is incomplete, it is often easier to do straight gravitational calculations rather than use conformal field theory (CFT) at weak coupling to learn about gravity.  For now the question of the fundamental nature of spacetime is too ambitious.  There are many simpler questions we can ask.  Which CFT configurations have gravity duals?  What types of curvature singularities are allowed?       

For strongly coupled gauge theories we want to use the gauge/gravity correspondence to explore the universal properties of matter.  The aim is to eventually do calculations in the strongly coupled regimes of quantum chromodynamics (QCD), relevant to describe for example the quark-gluon plasma produced in relativistic heavy ions collisions.  Using gravity at weak coupling to calculate hydrodynamic properties has been successful in yielding the entropy to shear viscosity ratio, see~ \cite{Policastro:2001yc}. Furthermore the authors conjectured this ratio to satisfy a universal lower bound, such that
\begin{eqnarray}
\frac{\eta}{s} \geq \frac{1}{4\pi}\,,
\end{eqnarray}
for any CFT with a gravity dual.

The key to doing this calculation was the observation that the long distance dynamics of any interacting quantum field theory near thermal equilibrium is well described by a relativistic fluid equation. In doing these calculations the real correspondence is between the fluid dynamics, an effective description of CFT, and gravity, an effective description of string theory.

In this review we will discuss recent techniques for calculating hydrodynamic properties using the fluid/gravity correspondence~\cite{Bhattacharyya:2008jc,Bhattacharyya:2008xc}.  The original methods for calculating dynamic properties of gauge fields using Minkowski correlators is reviewed in \cite{Son:2007vk}.   

%
\section{Review of the AdS/CFT correspondence\label{Sec:Review}}
%
The AdS/CFT correspondence we are interested in is the classic example between type IIB string theory on AdS$_5\times$S$^5$, a 10 dimensional theory of gravity, and $\mathcal{N} = 4$ supersymmetric Yang--Mills (SYM), a 4 dimensional gauge theory, see for example~\cite{Aharony:1999ti}.  The difference in dimension seems like a problem until one realizes that the extra dimensions on the gravity side become particle degrees of freedom on the gauge side. 

\begin{figure}[ht]
  \begin{center}
  \includegraphics{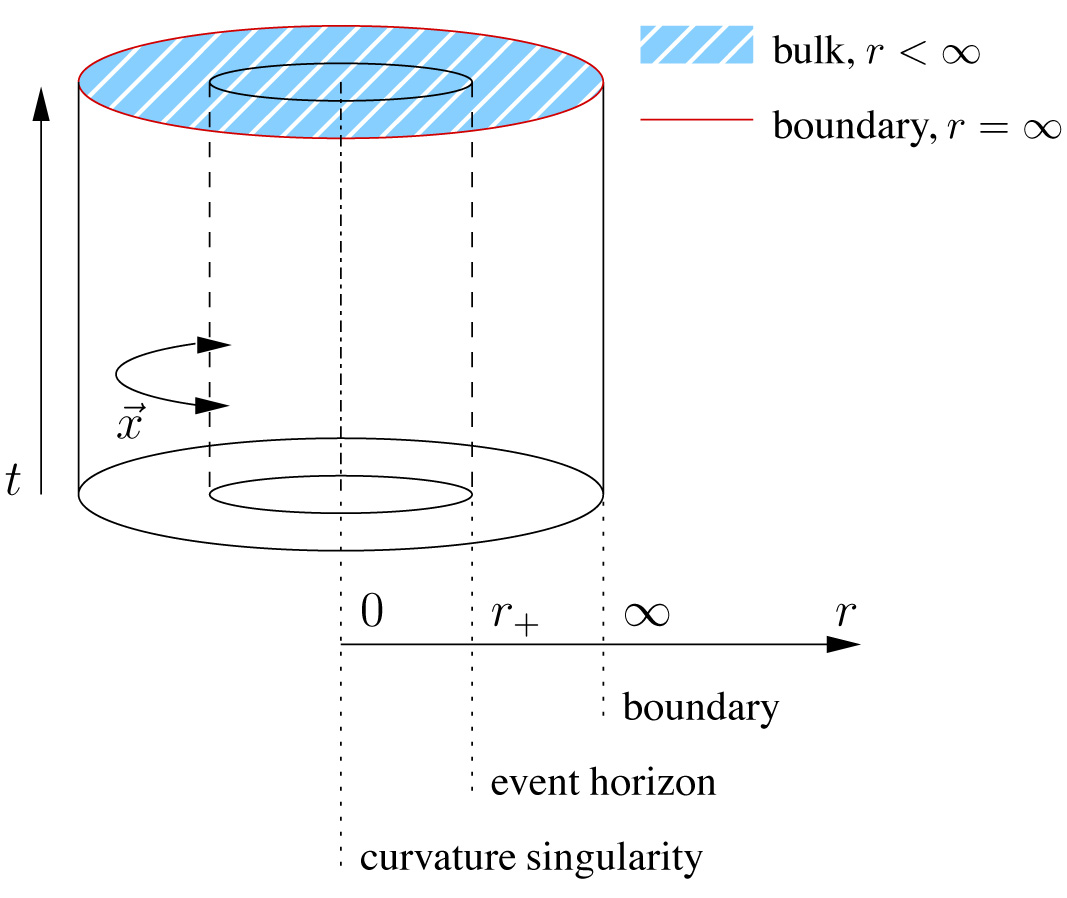}
  \caption{An illustration of where $\mathcal{N} = 4$ SYM lives on AdS$_5$ space.  The  4 dimensions of the gauge theory $(t,\mathbf{x})$ live on the boundary of the AdS$_5$ space at $r=\infty$.}
  \label{fig:ads}
  \end{center}
\end{figure}

In terms of the correspondence the gauge theory lives on the 4 dimensional boundary, located at $r=\infty$,  of the 5 dimensional AdS$_5$ space.  The dual gravity theory lives in the bulk of the AdS$_5$ space where $r<\infty$, see Figure~\ref{fig:ads}. The fundamental ideas behind the conjectured correspondence are outlined in Table~\ref{Summary1}.  Details of the dictionary between CFT operators on the boundary and field configurations in the bulk will be discussed later.  

Before using the correspondence for our calculations we must find an effective description for the theories that are dual to each other. Next we find static solutions to these effective theories from which it is possible to calculate static properties such as the entropy of strongly coupled SYM.  The real goal is to calculate dynamic properties by perturbing the static solution in one theory and seeing what that looks like in the other theory.  To do so we need to know what the perturbations in each theory look like. 

\begin{table}[ht]
\caption{\label{Summary1} Summary of the corresponding elements that appear in our duality.}
\begin{center}
  \begin{tabular}{ | p{5cm} | p{4.3cm} | p{4.3cm} | }   
    \hline
     & \textbf{bulk} & \textbf{boundary} \\
    \hline
    \textbf{AdS/CFT} & type IIB string theory on asymptotically AdS$_5\times$S$^5$  & $\mathcal{N}=4$ SYM on S$^3 \times\mathbb{R}^1$ or with a Poincar\'e patch $\mathbb{R}^3 \times\mathbb{R}^1$  \\
    \hline
    \textbf{effective description} & Einstein equation with cosmological constant & relativistic fluid dynamics  \\ 
    \hline
    \textbf{known static solutions} & black hole or black brane in AdS& static configuration of a perfect fluid \\
    \hline
    \textbf{perturbation} & non-uniformly evolving black branes& dissipative fluid flow \\
    \hline 
  \end{tabular}
\end{center}
\end{table}

\subsection{Regimes of validity\label{Sec:Validity}}
Since introducing why the duality exists is better left for another review, we will focus on the parameter matching that arises from the two theories being dual.  This is the simplest part of of the AdS/CFT dictionary. A more detailed discussion can be found in~\cite{Mateos:2007ay},  where a precise matching of fields and operators in the two theories is listed.

On the field theory side there are two parameters --- the number of colours $N$ (i.e., the rank of the gauge group  $SU(N)$) and the gauge coupling $g$.  When the number of colours is large, known as the planar limit, perturbation theory is controlled by the 't~Hooft coupling $\lambda = g^2N$.  On the string theory side the parameters are the string coupling $g_s$, the string length $l_s=\sqrt{\alpha'}$, and the radius $L$ of the AdS$_5$ space, which is proportional to $N^{1/4}$, where $N$ is the number of branes. The connection between the two originates from the double nature of D-branes: the gravitational and the gauge theoretical.

We will start by considering the gravity side.  A single Dp-brane has a tension that scales as $M\sim 1/g_s$, thus it is a nonperturbative object, and in the weak coupling limit can be treated as a rigid object. Vice-versa, at strong coupling Dp-branes become light and a large number, of order $N\sim 1/g_s$, are required to have a sizable gravitational effect. The Dp-brane is also charged under a $(p+1)\mbox{-form}$ potential with a charge $Q$ equal to its tension resulting in the BPS bound being saturated.\footnote{The BPS bound is in general $M\geq Q$.} In a supersymmetric theory saturation of the bound means that half of the supersymmetry is preserved and the other half is broken, but also that the configuration is stable. In our case this implies that we can stack an arbitrary number of Dp-branes and the configuration will remain stable. The mass of a black hole scales as $N^2\sim 1/g_s^2$ so stacking $N$ branes creates a background with a black brane equivalent of a black hole. The solution to the equations of motion of type IIB supergravity for a D3-brane is~\cite{Horowitz:1991cd,Gubser:1996de},
\begin{eqnarray}
\label{eq:extremalMetric} & ds^2 = H^{-1/2}(r)  \left[ -dt^2 + \sum_{i=1}^{3} (dx^i)^2 \right] + H^{1/2}(r) \left[  dr^2 + r^2 \, d\Omega_5^2 \right]  &\\ && \nonumber \\
& F_{(5)} = Q(\varepsilon + *\varepsilon)  \quad \mbox{and}  \quad \Phi = \mathrm{constant} , &
\end{eqnarray}     
where $H(r) = 1 + L^4/r^4$, the event horizon is at $r=r_+=0$ (compare with equation~(\ref{eq:NearExtremalD3Brane}) where $r_+\neq 0$), the volume form for the coordinates ($t,x^i,r$) is given by $\varepsilon$, and the Hodge star operator is denoted $*$.  The Ramond--Ramond $5$-form, $F_{(5)}$, is self dual and couples to the D3-brane, and the dilaton field $\Phi$ is constant. Since $g_s=e^\Phi$ we are free to choose any value for the string coupling. A more general, non-extremal, $r_+\neq 0$, version of the metric~(\ref{eq:extremalMetric}) is considered in section \ref{2ndLecture}.\footnote{If we take the near horizon limit, $r \sim r_+\ll L$, the geometry becomes that of AdS$_5\times$S$^5$ spacetime with radius $L$. From viewpoint of AdS$_5$ the $5$-form flux can be identified with the cosmological constant and can be neglected in the following discussion.}
It is possible to rewrite strings in terms of gravity  parameters $16\pi G = (2\pi)^7g_s^2l_s^8$ and by associating this black brane (black hole) with $N$ D3-branes we get
\begin{eqnarray}
\frac{L^4}{l_s^4} = 4\pi N\,.
\end{eqnarray}

Another relationship between parameters comes from the gauge theory created by the open strings attached to Dp-branes. The massless spectrum of the open strings on the Dp-brane is that of the maximally supersymmetric $(p+1)$ dimensional gauge theory with gauge group $SU(N)$ for $N$ stacked branes. In the case of D3-branes this is four dimensional $\mathcal{N}=4$ SYM. The effective action of the Dp-brane is the Dirac--Born--Infeld (DBI) action that, when expanded at first order in $\alpha'$, yields the usual Yang--Mills action. By identification of the coefficient of the gauge kinetic term we get the gauge coupling in terms of string theory parameters,
\begin{eqnarray}
g^2 &=& \frac{g_s}{l_s}(4\pi l_s)^{p-2}\, \\ \Rightarrow g^2 &=& 4\pi g_s\,, \ \textrm{for }p=3\,.
\end{eqnarray}
We are interested in $\mathcal{N} = 4$ SYM that is created by D3-branes, so  $p=3$, leaving a simple relationship between the gauge and string coupling constants.

We can rewrite these parameters as
\begin{eqnarray}
\frac{L^2}{\alpha'} \sim \sqrt{g_sN} \sim \sqrt{\lambda}\,.
\end{eqnarray}

The goal of this lecture series is to establish a correspondence between the gravitational limit of Type IIB string theory on AdS$_5$xS$^5$ space, and the hydrodynamic limit of the non-gravitational supersymmetric $\mathcal{N}=4$ Yang--Mills gauge theory defined on the conformal $4$-dimensional boundary of AdS$_5$.
For the gravitational description of string theory to be valid we require for the two dimensionless string parameters that
\begin{equation}
\frac{L}{l_s} = \frac{L}{\sqrt{\alpha'}} \gg 1 \quad \mbox{and} \quad g_s \ll 1,
\end{equation}
where the ratio between the curvature scale for the string background $L$ and the string length $l_s$ to be large (to suppress stringy effects), and simultaneously we assume the string coupling to be small (to further suppress quantum effects).
The equivalent of the gravitational limit in terms of fundamental parameters for the conformal field theory, the 't~Hooft coupling $\lambda$ and the Yang--Mills coupling $g$, is given by the following correspondence,
\begin{equation}
\left\{  \frac{L}{\sqrt{\alpha'}} , g_s  \right\} \quad \rightleftharpoons  \quad \left\{ \lambda=g^2 \, N , g^2  \right\}.
\end{equation}
Therefore the suppression of stringy and quantum effects on the boundary requires that
\begin{equation}
 \lambda \gg 1 \quad \mbox{and} \quad \frac{\lambda}{N} \ll 1 ,
\end{equation}
both $\lambda \rightarrow \infty$ and $N \rightarrow \infty$. This is the 't~Hooft limit with $\lambda\rightarrow \infty$.

Further, to obtain a hydrodynamical description for the boundary theory, we consider the local energy density of the conformal field theory such that we are able to thermodynamically associate a local notion of temperature $T$ and  mean free path $l_{\mathrm{mfp}} \sim 1/T$. The scale for the field fluctuations, $R$, has to be long wavelength in time and space. In other words, the scale of variations has to be large compared to the mean free path $l_\mathrm{mfp}\ll R$. In terms of the derivative expansion of the stress-energy tensor we want the first order term to be small compared to the zeroth order term,
\begin{equation}
\frac{1^\mathrm{st} \; \mbox{order}}{0^\mathrm{st}  \; \mbox{order}} \sim \frac{\eta \, \sigma^{\mu\nu}}{\rho \, u^{\mu}u^{\nu}} \sim \frac{\eta}{\rho \, R}
\equiv \frac{l_{\mathrm{mfp}}}{R} \sim \frac{1}{R \, T} \ll 1 ,
\,
\end{equation}
where we have assumed that $u^\mu \sim \mathcal{O}(1)$, and $\sigma^{\mu\nu}\sim 1/R$.

Using the parameter matching to write these in terms of AdS parameters we get,
\begin{eqnarray}
R \gg l_{\mathrm{mfp}} \Rightarrow r_+ \gg L\,.
\end{eqnarray}
So we see that the regime where the fluid is valid corresponds to a theory with large AdS black holes.

%
\section{Review of fluid dynamics\label{Sec:Review.Fluid.Dynamics}}
%
We are relatively familiar with the effective description of string theory on AdS backgrounds, that is classical (super)gravity, but less so with the effective description of SYM.  When first writing the fluid equations of motion we assume an ideal fluid - one that has no viscosity and no thermal conduction.  This means there is no energy dissipation and that the entropy of the fluid is constant. The effects we are interested in later are fundamentally dissipative so we will have to add corrections to these initially simplified equations of motion.  Thus we start with an adiabatic fluid and later we relax this constraint allowing the fluid parameters to vary slowly.

The standard description of fluid dynamics is given by the continuity and Euler equations.  The differential form of the continuity equation is obtained by realising that any change in the amount of fluid in a volume $V_0$ with a density $\rho$ must be accompanied by a fluid traveling at velocity $\mathbf{v}$ through the boundary of that volume, $\delta V_0$,
\begin{eqnarray}
\frac{\partial}{\partial t} \int_{V_0} \rho \, dV  &=& - \int_{\delta V_0} \rho \, \mathbf{v}\cdot \mathbf{dA} \\ &=& -\int_{V_0} \nabla \cdot (\rho \, \mathbf{v}) \, dV
\end{eqnarray}
where we have used Stokes theorem to rewrite the right hand side in terms of a volume integral. Since this is valid for any volume $V_0$ the equation of motion can be read off to be,
\begin{eqnarray}
\frac{\partial \rho}{\partial t} + \rho \nabla \cdot \mathbf{v} + \mathbf{v}\cdot\nabla \rho = 0 \,.
\end{eqnarray}
The Euler equation comes from Newton's law and relates the pressure $P$ of the fluid to its velocity
\begin{eqnarray}
-\nabla P = \rho\left(\frac{\partial \mathbf{v}}{\partial t} + (\mathbf{v}\cdot \nabla) \, \mathbf{v} \right)\,.
\end{eqnarray}

For our purposes it is more convenient to write these equations in a covariant manner by expressing the characteristics of the fluid in terms of a symmetric stress-energy tensor $T_{\mu\nu}$. Here $T_{00}$ represents the energy density, $T_{ii}$ the pressures in each direction, $T_{ij}$ the shear stresses, and $T_{0i}$ the momentum density.  Within this formalism the equations of motion can easily be written as
\begin{eqnarray}
\nabla^\mu T_{\mu\nu} = 0\,.
\end{eqnarray}
The stress-energy tensor describes a fluid of density $\rho(x^\mu)$, pressure $P(x^\mu)$, and fluid velocity $u^\nu(x^\mu)$, which is normalised to $u^\mu u_\mu = -1$.  The stress-energy tensor of any ideal fluid is not permitted to contain derivatives and thus must of the form of
\begin{eqnarray}
T_{\mu\nu} = (\textrm{scalar})u_\mu u_\nu + (\textrm{scalar}) g_{\mu\nu}\,.
\end{eqnarray}
If we further define a projection operator,
\begin{eqnarray}
P_{\mu\nu} = g_{\mu\nu} + u_\mu u_\nu \,,
\end{eqnarray}
that has the property
\begin{eqnarray}
P_{\mu\nu} u^\mu = 0\,.
\end{eqnarray}
We can make use of dimensionality arguments to find the explicit expression for the stress-energy tensor
\begin{eqnarray}
T_{\mu\nu} = \rho \, u_\mu u_\nu + P P_{\mu\nu}\,.
\end{eqnarray}

\vspace{5mm}\noindent{\bf  Exercise:}. {\it
Show that the relativistic versions of the ideal fluid equations can be obtained by replacing $T_{\mu\nu} = \rho \, u_\mu u_\nu + P P_{\mu\nu}$ into the equation of motion $\nabla^\mu T_{\mu\nu} = 0$.
}\vspace{5mm}

Ultimately, we are interested in calculating dissipative properties of conformal fluids.  The stress-energy tensor up to zeroth order, being void of dissipation, captures none of these. We need to look at the first order dissipative corrections to the stress-energy tensor.

We want to construct the most general $n-$derivative dissipative correction order by order.  It should be proportional to a single derivative of the fluid velocity $\nabla_\mu u_\nu$.  This tensor can be decomposed into irreducible representations, separating into components parallel or orthogonal to $u_\mu$
\begin{eqnarray}
\nabla_\mu u_\nu = -a_\mu u_\nu + \sigma_{\mu\nu} + \omega_{\mu\nu} + \frac{1}{3}\theta P_{\mu\nu}\,,
\end{eqnarray}  
where the trace is fully contained in the last term, $\theta = \nabla_\mu u^\mu$, and the first three terms are left traceless.  The first term contains the acceleration, $a^\mu = u^\nu\nabla_\nu u^\mu$, that is orthogonal to the projection tensor $P_{\mu\nu}$.  The second term is the shear,
\begin{eqnarray}
\sigma^{\mu\nu} = \nabla^{(\mu}u^{\nu)} + u^{(\mu}a^{\nu)} - \frac{1}{3}\theta P^{\mu\nu}\,,
\end{eqnarray}
which is symmetric and traceless by construction, and is orthogonal to $u^\mu$. The vorticity,
\begin{eqnarray}
\omega^{\mu\nu} = \nabla^{[\mu}u^{\nu]} + u^{[\mu}a^{\nu]} \,,
\end{eqnarray}
is also orthogonal to $u^\mu$ and is antisymmetric by construction.

Now that each irreducible representation has been identified with a physical quantity we can find the constants of proportionality to make it a correction to the stress-energy tensor.  We find
\begin{eqnarray} \label{equ:stress.tensor}
T^{\mu\nu}_{\mathrm{dissip}} = -\zeta\theta P_{\mu\nu} - 2\eta\sigma^{\mu\nu} + 2 q^{(\mu}u^{\nu)}\,,
\end{eqnarray}
where $\zeta$ is the bulk viscosity, $\eta$ is the shear viscosity, and the last term is constructed from the vorticity and represents the heat dissipation $q=-\kappa P^{\mu\nu}(\delta_\mu T + a_\nu T)$.

The appearance of a the dissipation term can also be quantified by looking at the derivative of the entropy current $J_\mu^{(s)} = su_\mu$, where $s$ is the entropy density.  At zeroth order we assume an adiabatic fluid which means the entropy is constant.  To obtain this we set all derivatives of $u_\mu$ to zero.  Clearly then the entropy current is conserved,
\begin{eqnarray}
\nabla^\mu J_\mu^{(s)} = 0\,.
\end{eqnarray}
When dissipative terms are introduced there are now non-zero derivatives of $u^\mu$ and the entropy current is no longer conserved,
\begin{eqnarray}
T\nabla^\mu J_\mu^{(s)} &=& \frac{q_\mu q^\nu}{\kappa T} + \zeta\theta^2 + 2\eta\sigma_{\mu\nu}\sigma^{\mu\nu} \\ &\geq& 0\,.
\end{eqnarray}
We can see that because everything is squared, the correction is always positive, and the entropy always increases with dissipation. 

For the AdS/CFT correspondence it is necessary to constrain $T_{\mu\nu}$ to represent conformal fluid equations. Before we proceed we would like to briefly recall the definition for conformal invariance.

\vspace{5mm}\noindent{\bf Conformal invariance}. {\it
Consider a rescaling of the metric field tensor by some conformal factor, $g_{\mu\nu}=e^{2\, \phi} \tilde g_{\mu\nu}$, involving a scalar field $\phi$. Further consider a field $\psi$ satisfying the field equations $H[\psi, g_{\mu\nu}]=0$, depending on $\psi$ and $g_{\mu\nu}$. $H$ is said to be conformally invariant, if and only if we can find a conformal weight $s$ of $\psi$, where $s\in \mathbb{R}$, such that $\tilde\psi$ and $\tilde g_{\mu\nu}$, where $\psi=e^{s\, \phi} \tilde\psi$ and $g_{\mu\nu}=e^{2\, \phi} \tilde{g}_{\mu\nu}$, also satisfy the field equations, $H[\tilde\psi, \tilde g_{\mu\nu}]=0$.

For example, the relativistic wave equation for massless scalar field minimally coupled to the gravitational field $g_{\mu\nu}$,
\begin{equation}
\partial_{\mu} \left( \sqrt{-g} \, g^{\mu\nu} \, \partial_{\nu} \psi \right) = 0 ,
\end{equation}
is conformally invariant in $d=2$, where $s=0$. However, we are able to extend the equation of motion for $\psi$ such that the fields couple conformally to the metric tensor,
\begin{equation}
\partial_{\mu} \left( \sqrt{-g} \, g^{\mu\nu} \, \partial_{\nu} \psi \right)  + \frac{d-2}{4\, (d-1)} R \, \psi= 0 ,
\end{equation}
for $d>1$ and with $s=1-d/2$.
}\vspace{5mm}

In order to have the conformal invariance (i.e., invariance under rescaling of $T^{\mu\nu}=e^{s\phi} \tilde{T}^{\mu\nu}$ and $g_{\mu\nu}=e^{2\phi}\tilde{g}_{\mu\nu}$) of the relativistic Navier--Stokes equation, $\nabla_\mu T^{\mu\nu}=0$, it is necessary to demand
\begin{equation}
{T_{\mu}}^{\mu} = 0  \quad \mbox{and}\quad s=-d-2 \quad \mbox{(here}\, d=4 \,\, \mbox{thus} \,\, s=-6 \mbox{)}.
\end{equation}
Thus conformal invariance implies Weyl symmetry, that is the trace of $T_{\mu\nu}$ must vanish, and the equation of state for a perfect fluid in $d-dimensions$ reduces to
\begin{equation}
P = \frac{\rho}{d-1}.
\end{equation}
Altogether, to first order, in terms of $T(x^\mu)$ and $u^\nu(x^\mu)$ we get
\begin{equation}
T^{\mu\nu} = (\pi T)^4 \, \left( \eta^{\mu\nu} + 4 \, u^{\mu} u^{\nu} \right) - 2 \, (\pi T)^3 \, \sigma^{\mu\nu}  .
\end{equation}
With this equation we are able to read off the entropy to shear viscosity defined earlier on 
\begin{equation}
\frac{\eta}{s} = \frac{\pi^3 T^3}{4 \pi^4 T^3} = \frac{1}{4 \pi}.
\end{equation}
This can easily be seen as the shear viscosity $\eta$ is given by the factor in front of the shear tensor $\sigma_{\mu\nu}$, compare with equation~(\ref{equ:stress.tensor}), and the entropy is defined by $s=\partial T^{00} / \partial T$ where $T^{00}=(\pi T)^4$.
Please note that the variable $s$ has previously been used in a different context where it expressed the conformal weight.

%
\section{\label{2ndLecture}Schwarzschild AdS$_{5} \times$S$^5$ black hole (black branes)}
\begin{figure}[ht]
\begin{center}
\includegraphics{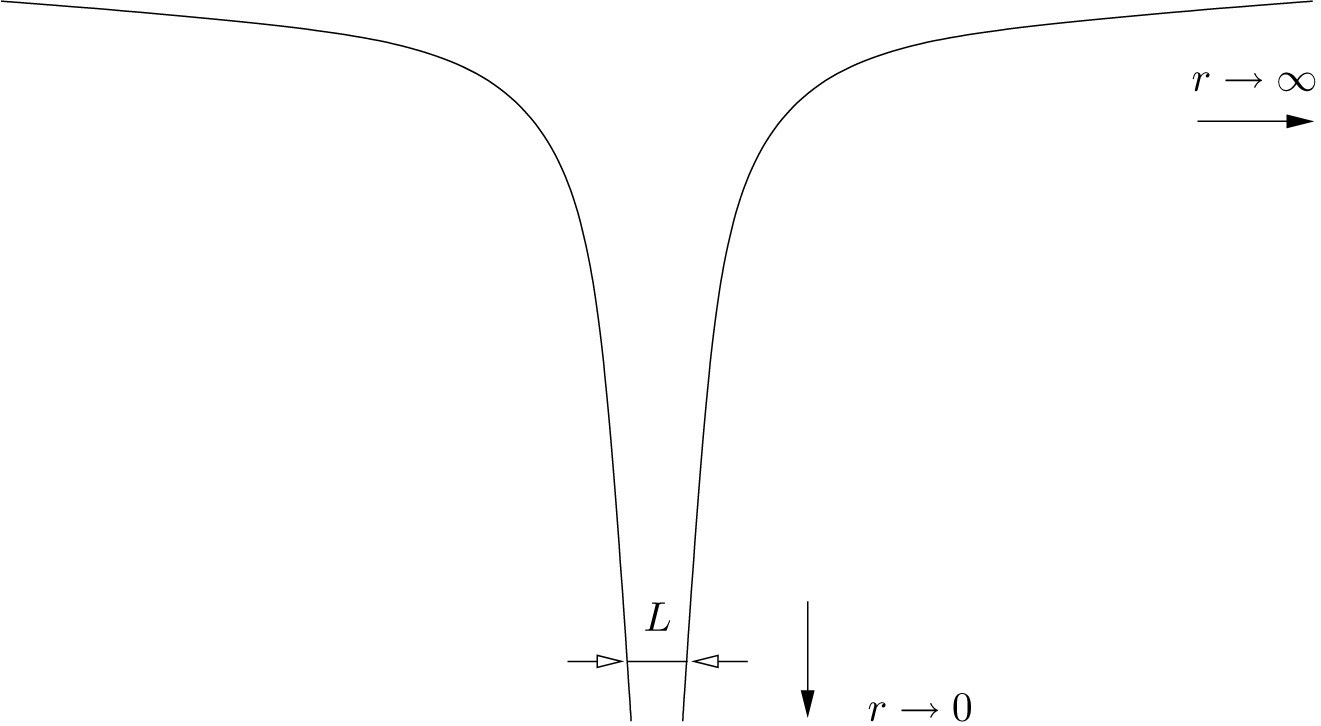}
\caption[Near horizon limit.]  {\label{fig:SF1}
The figure illustrates the various limits. For $r \sim r_+\ll L$ we are in the near horizon limit, while for $r/L \rightarrow \infty$ we are asymptotically approaching flat spacetime.}
\end{center}
\end{figure}

%
We will study the gravity side of the correspondence provided by the underlying string theory.
A static solution for Einstein equations with negative cosmological constant in the bulk for $N$ near extremal D3 branes is given by
\begin{equation} \label{eq:NearExtremalD3Brane}
ds^2 = H^{-1/2}(r)  \left[ -f(r) \, dt^2 + \sum_{i=1}^{3} (dx^i)^2 \right] + H^{1/2}(r) \left[ f^{-1}(r) \, dr^2 + r^2 \, d\Omega_5^2 \right] ,
\end{equation}
where we have introduced the following parameters
\begin{equation}
H(r) = 1 + \frac{L^4}{r^4} \quad \mbox{and} \quad f(r)=1-\frac{r_+^4}{r^4}.
\end{equation}
Notice that for $r_+ = 0$ the extremal D3 brane, see equation~(\ref{eq:extremalMetric}), has its horizon at $r=0$, see Figure~\ref{fig:SF1}, and at the near horizon limit, where $r \sim r_+\ll L$ and thus $H(r)=L^4 / r^4$, the metric simplifies to
\begin{equation}
ds^2 = \frac{r^2}{L^2} \left[ -\left( 1- \frac{r_+}{r^4} \right) dt^2  + \sum_{i=1}^{3} (dx^i)^2  \right]
+ \frac{L^2}{r^2} \left( 1-\frac{r_+^4}{r^4}  \right)^{-1} dr^2 + L^2 \, d\Omega_5^2 .
\end{equation}
%
%

\begin{figure}[ht]
  \begin{center}
  \includegraphics[width=12cm]{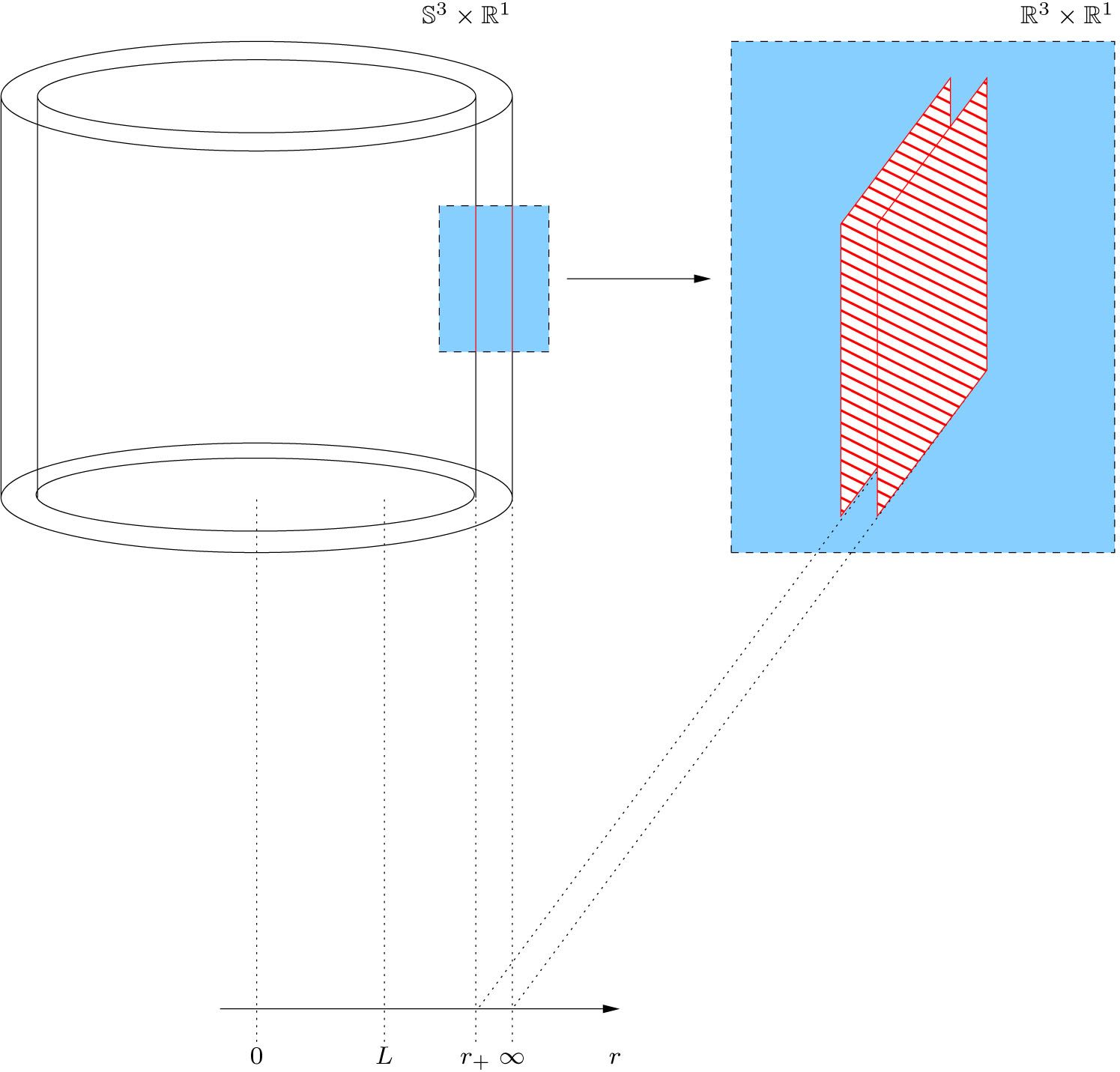}
  \caption{Planar limit, $r\sim r_+\gg L$.   \label{fig:SF2}}
  \end{center}
\end{figure}

It is also interesting to compare the near-extremal D3-brane with the global Schwarzschild AdS$_5$ black hole,
\begin{equation}
ds^2=-h(r) \, dt^2 + \frac{dr^2 }{h(r)}  + r^2 \, d\Omega_3^2 + L^2 \, d\Omega_5^2 \, ,
\end{equation}
where
\begin{equation}
h(r)=\frac{r^2}{L^2} + 1 - \frac{r_0^2}{r^2} \equiv \frac{r^2}{L^2} + 1 - \frac{r_+^2}{r^2} \left( \frac{r_+^2}{L^2} + 1 \right).
\end{equation}
In the planar limit, $r_+ \gg L$, thus
\begin{equation}
h(r) \rightarrow \frac{r^2}{L^2} - \frac{r_0^2}{r^2} \quad \mbox{and} \quad r_0^2 \rightarrow \frac{r_+^4}{L^2},
\end{equation}
and the $3$-sphere decompactifies,
\begin{equation}
r^2 \, d\Omega_3^2 \rightarrow r^2 \, \frac{dx_i \, dx^i}{L^2}.
\end{equation}
Substituting this in we get the our near horizon metric.  So the planar limit $r_+ \gg L$ of the global Schw-AdS$_5$ metric gives the near horizon limit of the string theory metric for $N$ near extremal black branes, see Figure~\ref{fig:SF2}.
%
%

\subsection{Key features of AdS$_5\times$S$^5$\label{Sec:Causal.Structure}}
Before continuing we would like to summarize the key features of the gravitational field in the bulk.
There exists a scaling symmetry in the planar limit,
\begin{equation}
t \rightarrow \lambda \, t, \quad x^i \rightarrow \lambda \, x^i, \quad \mbox{and} \quad r \rightarrow \frac{r}{\lambda} \Rightarrow r_+ \rightarrow \frac{r_+}{\lambda} ,
\end{equation}
such that $ds^2 \rightarrow ds^2$ with $r_+$ rescaled.
Therefore it can be shown that the Hawking temperature~\cite{Hawking:1974rv,Hawking:1974sw}, given by
\begin{equation}
T_{\mathrm{BH}} = \frac{h'(r_+)}{4 \, \pi} = \frac{2\, r_+^2 + L^2}{2 \pi \, r_+ \, L^2}, \quad  T_\mathrm{BH} = \frac{r_+}{\pi \, L^2} \rightarrow
\frac{T_\mathrm{BH}}{\lambda},
\end{equation}
is also part of the scaling symmetry.

For the causal structure, see the Penrose diagram in Figure~\ref{SubFig:PenroseAdS}, we are interested in the asymptotic behaviour, the existence of horizons and curvature singularities.
As $r\rightarrow \infty$ we get for the line element
\begin{equation}
ds^2 \rightarrow \frac{r^2}{L^2} \, \left( -dt^2 + d\vec{x}^2 \right) + \frac{L^2}{r^2} dr^2 + L^2 \, d\Omega_5^2,
\end{equation}
corresponding to an asymptotic AdS$_5$xS$^5$ spacetime.
There exists a Killing-vector field $\partial_t$ with norm given by $g_{tt}$, and at $r=r_+$ the norm of Killing-vector field vanishes, thus there exists a Killing horizon at $r_+$.
Finally, it can be shown by studying the behaviour of the diffeomorphism invariant quantity $R^{abcd}R_{abcd}$, the square of the Riemann tensor, that at $r=0$ we are dealing with a curvature singularity, where $R^{abcd}R_{abcd}\rightarrow \infty$.

%
\begin{figure*}[!htb]
\begin{center}
\mbox{
\subfigure[Schwarzschild black hole in asymptotically flat spacetime. \label{SubFig:PenroseFlat}]{\includegraphics{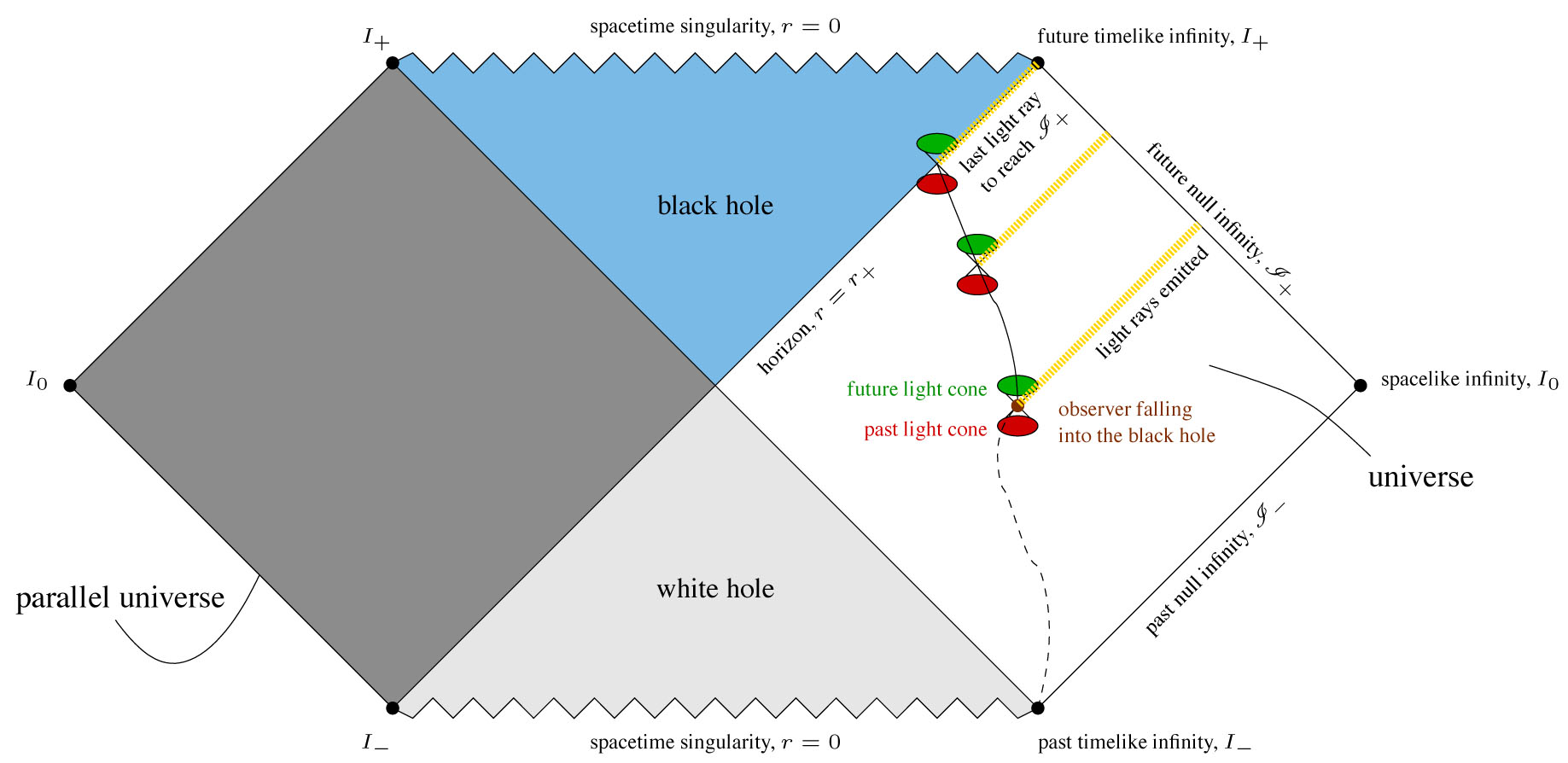}}
\hspace{0mm}
}
\mbox{
\subfigure[Schwarzschild black hole in asymptotically AdS spacetime. \label{SubFig:PenroseAdS}]{\includegraphics{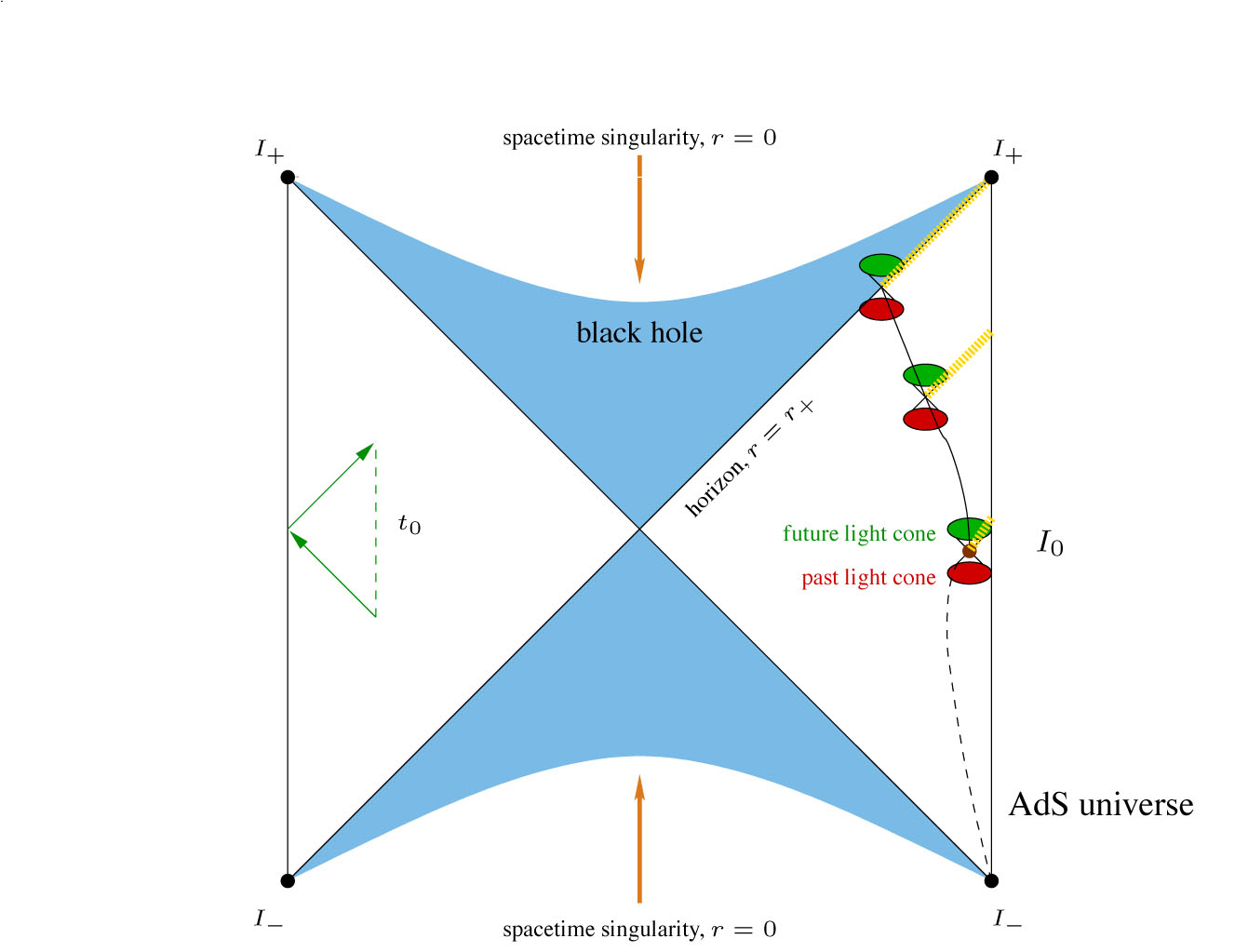}}
  }
\caption[PenroseDiagrams  \label{SubFig:PenroseAdS}]{The figures illustrate the Penrose diagrams for black holes in spacetimes with different asymptotic behaviour. Here light rays propagate along $45^\circ$ lines, see yellow dashed lines as emitted from an observer falling into the black hole.}\label{Fig:Penrose}.
\end{center}
\end{figure*}
%

It is also insightful to directly compare the causal structure for a Schwarzschild black hole in two different spacetimes --- asymptotic Minkowski/flat and AdS --- using Penrose diagrams\footnote{Penrose diagrams are plots of conformally transformed spacetimes. The conformal factor has been chosen such that the entire (infinite) spacetime is mapped onto a finite region. These diagrams are especially useful to study causal relations between any two points in the ``original'' spacetime, as conformal transformations maintain the light-cone structure.}.

In Figure~\ref{SubFig:PenroseFlat} we plotted the worldline for an observer in an asymptotically flat spacetime  that starts at past timelike infinity $I_-$. Once the observer crosses the horizon at $r=r_+$ no light rays can be transmitted to future null infinity $\mathscr{I}_+$ (i.e., to any other observer outside the black hole) and the infalling observer inevitably falls towards the spacetime singularity at $r=0$. Consequently no information can leave the black hole. This is the opposite of a white hole where no information can enter. The white hole is obtained from \textit{mathematically} extending the spacetime solution. To make the spacetime diagram more accessible we indicate future and past null infinity as $\mathscr{I}_\pm$, future and past timelike infinty as $I_\pm$, and spacelike infinity $I_0$. These indicate respectively where lightlike, timelike, and spacelike worldlines start and end.

The global structure for a Schwarzschild black hole in an asymptotic AdS spacetime is shown in Figure~\ref{SubFig:PenroseAdS}. Here $\mathscr{I}_\pm=I_0$ and spatial infinity is now an extended line rather than a single point. A consequence of the extended spacelike infinity is that a single light ray can reach infinity, bounce back, and return to its origin in a finite time $t_0$, see green lines in Figure~\ref{SubFig:PenroseAdS}.

From now on we will drop the term, $L^2d\Omega_5^2$, in the metric on $S^5$ and use
\begin{equation}
        ds^2=\frac{r^2}{L^2}\left[-\left(1-\frac{r_+^4}{r^4}\right)dt^2+\sum_{i=1}^3 (dx^i)^2\right]+\frac{L^2}{r^2}\left(1-\frac{r_+^4}{r^4}\right)^{-1}dr^2\,.
        \label{eq:planar-shw-ads5}
\end{equation}
The extra degrees of freedom corresponding to fluctuations around the S$^5$ are dual to operators on the CFT side that we are not currently interested in. For the metric (\ref{eq:planar-shw-ads5}) the boundary lies at $r=\infty$, the event horizon lies at $r=r_+$, there is a curvature singularity at $r=0$, and the black hole temperature, which corresponds to the thermal temperature of the gauge theory, is $T_{\mathrm{BH}}=\frac{r_+}{\pi L^2}$, thus all the key features are preserved.  

\subsection{Boosted black branes\label{Sec:Boosted}}
One way to avoid the coordinate singularity at $r=r_+$ is to use  ``ingoing'' Eddington-Finkelstein coordinates, $(V,x^i,r, \cdots)$, where
\begin{equation}
    V=t+r_* \quad \mbox{and} \quad
    dr_*= \frac{dr}{\frac{r^2}{L^2}\left(1-\frac{r_+^4}{r^4}\right)}\,.
    \label{eq:new coord}
\end{equation}
Under the new coordinates $(V,x^i,r)$ the metric of the planar Schw-AdS$_5$ black hole (\ref{eq:planar-shw-ads5}) can be rewritten as
\begin{equation}
    ds^2=-\frac{r^2}{L^2}\left(1-\frac{r_+^4}{r^4}\right)dV^2+2dVdr+\frac{r^2}{L^2}\sum_{i=1}^{3}(dx^i)^2\,.
    \label{eq:newads5}
\end{equation}
We now find that $g_{\mu\nu}$ and $g^{\mu\nu}$ remain finite $\forall\ r>0$ and we have an ingoing null geodesic: $x^{\mu}= 0$, and  $r= \mbox{constant}$.

Covariantizing $x^\mu=(V,x^i)$ with respect to the boundary directions we get
\begin{eqnarray}
    V=-u_\mu x^\mu & \Rightarrow & dV=-u_\mu dx^\mu \Rightarrow dV^2=u_\mu u_\nu dx^\mu dx^\nu \\
    x^i=P_\mu^i x^\mu  & \Rightarrow & dx_idx^i=P_{\mu i}P_\nu ^i dx^\mu dx^\nu=P_{\mu\nu}dx^\mu dx^\nu\,,
\end{eqnarray}
where $u^\mu=(1,0,0,0), u_\mu=(-1,0,0,0)$.
Substituting them into the metric of the planar Schw-AdS$_5$ black hole we have
\begin{eqnarray}\label{eq:boost}
    ds^2&=&-\frac{r^2}{L^2}\left(1-\frac{r_+^4}{r^4}\right) u_\mu x_\nu dx^\mu dx^\nu-2u_\mu dx^\mu dr+\frac{r^2}{L^2} P_{\mu\nu}dx^\mu dx^\nu \nonumber \\
        &=& -2u_\mu dx^\mu dr+\frac{r^2}{L^2}\left(\eta_{\mu\nu}+\frac{r_+^4}{r^4}u_\mu u_\nu\right)dx^\mu dx^\nu\,.
\end{eqnarray}
The metric (\ref{eq:boost}) is called as the boosted uniform black hole.

\section{Inducing a stress-energy tensor with a metric}

Now we consider the boundary stress-energy tensor $T_{\mu\nu}$ for the asymptotical AdS geometry, where we will use the unboosted metric,
\begin{equation} \label{eq:initialMetric}
        ds^2=\frac{r^2}{L^2}\left[-\left(1-\frac{r_+^4}{r^4}\right)dt^2+\sum_{i=1}^3 (dx^i)^2\right]+\frac{L^2}{r^2}\left[1-\frac{r_+^4}{r^4}\right]^{-1}dr^2\,.
\end{equation}
We have already established what the stress-energy tensor looks like in terms of dynamical fluid variables and that it lives on the boundary of the AdS$_5$ space.  To do calculations using the correspondence we are interested in what the gravitational stress-energy tensor on the boundary looks like.  Knowing this we are able to calculate the stress-energy tensor.

The correct prescription for determining the stress-energy tensor on the boundary given a metric in any dimension is found in~\cite{Balasubramanian:1999re}.  The method requires one to consider the boundary surface $\partial M_n$ of the $n$ dimensional bulk $ M_n$. To do this we want to foliate the spacetime such that the slices are parallel to the boundary and calculate the stress-energy tensor for this foliation, see for example~\cite{Poisson:2004aa}. In particular we are interested in its behaviour on the boundary. We would like to point out that this prescription gives a unique stress-energy tensor on the boundary implying that the stress-energy tensor induced for a given metric is unique~\cite{Balasubramanian:1999re}.

The first thing we do is find the stress-energy tensor in the bulk.  The gravitational action with cosmological constant $\Lambda$ in $n+1$ dimensions is,
\begin{eqnarray}
S = -\frac{1}{16\pi G}\int_M d^{n+1}x \sqrt{-g} (R-2\Lambda) - \frac{1}{8\pi G}\int_{\partial M} d^{n}x \sqrt{-\gamma} K + \frac{1}{8\pi G}S_{ct}(\gamma_{\mu\nu}) \,.
\end{eqnarray}
The first term is the bulk action whose solutions $\delta S_{M_{n+1}} = 0$ for AdS$_{n+1}$ with curvature $L$ gives $\Lambda = -\frac{n(n-1)}{2L^2}$, the cosmological constant.  The second term is a surface term that contains the extrinsic curvature $K_{\mu\nu} = -{\gamma_\mu}^\sigma \nabla_\sigma n_\nu$, which is a Lie derivative of $\gamma_{\mu\nu}$ pointing in the direction of $n_\mu$.  The action also contains counter terms to cancel the divergences due to the infinite volume of AdS$_5$ and obtain a finite stress-energy tensor $T_{\mu\nu}$. The form of these counter terms depend on the number of dimensions. For the AdS$_5$ case the counter terms are $\mathcal{L}_{ct} = -\frac{3}{L} \sqrt{-\gamma}(1 - \frac{L^2}{12} R(\gamma)) $.

Here we are concerned only with the AdS$_5$ result. The general result in other dimensions can be found in~\cite{Balasubramanian:1999re}. The boundary stress-energy tensor is
\begin{eqnarray} \label{Eq:STB}
T_{\mu\nu} = \frac{2}{\sqrt{-\gamma}} \frac{\delta S_{cl}}{\delta \gamma^{\mu\nu}}=  \frac{1}{8\pi G} \left[K_{\mu\nu} - K \gamma_{\mu\nu}  -\frac{3}{L} \gamma_{\mu\nu}  - \frac{L}{2}G_{\mu\nu}   \right] .
\end{eqnarray}
This is obtained in the standard manner by varying the action with respect to the boundary metric $\gamma_{\mu\nu}$. Note that the last two terms came from the counter term in the Lagrangian. 

The problem is then reduced to finding the extrinsic curvature. We must choose a curve to slice along.  Since we are looking at how slices approach the boundary the natural choice is along the radial coordinate of the AdS$_5$ space, 
\begin{eqnarray}
d\tilde{r} = \sqrt{g_{rr}}dr\,,
\end{eqnarray}
for any constant $t$ and $ x_i$.  With this slicing the metric decomposes into,
\begin{eqnarray}
ds^2 &=& N^2 dr^2 + \gamma_{\mu\nu}(dx^\mu + V^\mu dr)(dx^\nu+ V^\nu dr)\\
&=& N^2 dr^2 + \gamma_{\mu\nu}dx^\mu dx^\nu\,,
\end{eqnarray}
where $\gamma_{\mu\nu}$ is the induced metric given by,
\begin{eqnarray}
\gamma_{\mu\nu} = g_{\alpha\beta} \, e^\alpha_\mu \,e^\beta_\nu \quad \mbox{where} \quad e^\alpha_\mu = \left( \frac{\partial x^\alpha}{\partial \tilde{x}^\mu} \right)\,,
\end{eqnarray}
and $\mu,\nu$ run from $0$ to $3$, $\alpha,\beta$ run from $0$ to $4$.  The initial metric~(\ref{eq:initialMetric}) has no off-diagonal elements such that $V^\mu = 0$. The extrinsic curvature is then given as,
\begin{eqnarray}
K_{\mu\nu} &=& - \frac{1}{2N}n^\sigma \partial_\sigma \gamma_{\mu\nu} + D_\mu V_\nu + D_\nu V_\mu \\
&=&  -\frac{1}{2N} n^\sigma \partial_\sigma \gamma_{\mu\nu}\,.
\end{eqnarray}
We are left then with finding the normal vector $n^\mu$ and the form of the induced metric $\gamma_{\mu\nu}$. The vector normal to the hypersurface is given by
\begin{eqnarray}
n_\alpha &=& N \, \partial_\alpha \tilde{r}\\
&=& N\sqrt{g_{rr}}{\delta_\alpha}^r\,.
\end{eqnarray}
Raising the index gives
\begin{eqnarray}
n^\alpha &=& n_\beta g^{\alpha\beta} = N\frac{1}{\sqrt{g_{rr}}}{\delta_r}^\alpha \\
&=& N \sqrt{\frac{r^2}{L^2}\left(1 -\frac{r_+^4}{r^4} \right)} \, {\delta_r}^\alpha\,.
\end{eqnarray}
Due to the simple diagonal structure and the exclusive dependence of the metric on $r$ we can set $\tilde{t}=t$ and $\tilde{x}_i = x_i$, thus all
\begin{eqnarray}
e^\alpha_\mu = 1\,.
\end{eqnarray}
The induced metric is given by
\begin{equation}
d \gamma^2=\frac{r^2}{L^2}\left[-\left(1-\frac{r_+^4}{r^4}\right)dt^2+\sum_{i=1}^3 (dx^i)^2\right],
\end{equation}
where $r=r(\tilde{r})$.

The Einstein tensor $G_{\mu\nu} (\gamma)$ vanishes at the boundary and using the previous results we can find $T_{\mu\nu}$, see equation~(\ref{Eq:STB}), for a metric $g_{\alpha\beta}$ corresponding to planar Schw-AdS$_5$.  We first ensure that the divergent terms cancel out and then it can been shown that,
\begin{eqnarray}
T^{tt} &=& \frac{3}{16\pi G} \frac{Lr_+^4}{r^6} + \mathcal{O}\left(\frac{1}{r^{10}}\right) \\ T^{ii} &=& \frac{1}{16 \pi G} \frac{r_+^4}{r^6} + ... \, .
\end{eqnarray}
Altogether the stress-energy tensor on the boundary is
\begin{eqnarray}
T_{\partial M}^{\mu\nu}= \lim_{r \rightarrow  \infty } r^6 \, T^{\mu\nu} = \frac{1}{16\pi G}\left[(\pi T)^4(\eta^{\mu\nu} + 4u^\mu u^\nu) \right]\,,
\end{eqnarray}
where we can rewrite $\eta^{\mu\nu} + 4u^\mu u^\nu=P^{\mu\nu} + 3u^\mu u^\nu$.  We now have the form of the stress-energy tensor on the gravity side that is induced on the boundary.

%
\section{Introducing dynamics and non-uniformity\label{Sec:Dynamics.Non-Uniformly}}
%
Without loss of generality we can set
\begin{equation}
    L=1 \, , \quad \quad  f(r)=1-\frac{1}{r^4}\, \quad \mbox{and} \quad b=\frac{1}{\pi T}=\frac{1}{r_+} \, ,
\end{equation}
such that
\begin{equation}
f(b\,r)=1-\frac{r_+^4}{r^4} \,.
\end{equation}
At zeroth order the boosted black brane metric corresponds to the stress-energy tensor of a perfect fluid. A summary of the current picture can be found in Table~\ref{Summary2}, with the stress-energy tensor induced from the metric as explained in the previous section.

\begin{table}[htdp]
\caption{\label{Summary2} Summary for the relevant parameters in the bulk ($5$ dimensional metric tensor) and on the boundary ($4$ dimensional stress-energy tensor) and the corresponding equations of motion the Einstein equations in the bulk and the relativisitic hydrodynamic equations for the boundary.}
\begin{center}
\begin{tabular}{|c|c|}
\hline
Bulk & Boundary\\
\hline
$E_{MN}=R_{MN}-\frac{1}{2}Rg_{MN}+\Lambda g_{MN}=0$ & $\partial_\mu T^{\mu\nu}=0$\\
\hline
$ds^2=-2u_\mu dx^\mu dr+r^2\left[\eta_{\mu\nu}+(1-f(br))u_\mu u_\nu\right] dx^\mu dx^\nu$ & $T^{\mu\nu}=\frac{1}{b^4}(4u^\mu u^\nu+\eta^{\mu\nu})$\\
\hline
\end{tabular}
\end{center}
\end{table}

We then look for solutions of Einstein's equations where the parameters $\{b,u^i\}$ are promoted to \textit{slowly} varying functions of the boundary coordinates $x^\mu$. Henceforth we shall call the metric with the parameters promoted to functions $g^{(0)}$,

\begin{equation}
g|_{b\rightarrow b(x)\,,u^i \rightarrow u^i(x)}=g^{(0)}[b(x),u^i(x)]\,.
\end{equation}
But $g^{(0)}$ is no longer a solution of the Einstein equations for arbitrary $b(x)$, $u^i(x)$. Nevertheless it has two nice features; it is manifestly regular (i.e., non-singular) for every positive $r$ and for slowly varying $\{b(x),u^i(x)\}$ we expect it to be a good approximation to the true solution since locally in $x^\mu$ it can be ``tubewise" well approximated by a boosted black brane.

For a fluid in local thermal equilibrium with typical fluctuations $R$ much larger than the scale of inverse temperature we can expand the solution in a series expansion in $\frac{b}{R}\sim \frac{1}{TR}\equiv \epsilon\ll 1$. Having order $\epsilon^n$ then corresponds to $n$ boundary derivatives. Inserting the ansatz $g^{(0)}$ into Einstein's equations we get

\begin{equation}
E_{MN}\left[g^{(0)}\right]=\mathcal{O}(\epsilon)\neq 0 \, ,
\end{equation}
where $\mathcal{O}(\epsilon)$ correspond to terms with derivatives and the linear term has vanished because it is a solution of the equations. The approach is to expand the metric in $\epsilon$

\begin{equation}\label{perturbansatz}
g_{MN}=g^{(0)}_{MN} + \epsilon\, g^{(1)}_{MN} + \epsilon^2\, g^{(2)}_{MN}+\ldots \, ,
\end{equation}
where the $g^{(i)}_{MN}$ depend on $\{b(x),u^i(x)\}$ and are correction terms such that $g_{MN}$ solves $E_{MN}=0$ to a given order in $\epsilon$. This will be possible only if $\{b(x),u^i(x)\}$ satisfy certain equations of motion (namely $\partial_\mu T^{\mu\nu}=0$) that are corrected order by order in $\epsilon$. Consequently we must also correct $\{b(x),u^i(x)\}$ order by order in $\epsilon$, so we expand them as

\begin{eqnarray}
    b=b^{(0)}+\epsilon\, b^{(1)}+\ldots \\ u_i=u_i^{(0)}+\epsilon\, u_i^{(1)}+\ldots
\end{eqnarray}
with constant $\{b^{(i)},u_i^{(j)}\}$.

\subsection{General structure of perturbation theory\label{pertstruc}}
Now we can solve for $g^{(i)}$ iteratively. Let us imagine that we have solved the perturbation theory to the
$(n-1)^{{\rm th}}$ order (i.e., we have determined $g^{(m)}$ for $m\leq n-1$) and we
have determined the functions $u_i^{(m)}$ and $b^{(m)}$ for $m \leq n-2$.
Inserting the expansion~(\ref{perturbansatz}) into the Einstein equation as given in Table~\ref{Summary2},
and extracting the coefficient of $\epsilon^n$,  we
obtain an equation of the form
\begin{equation} \label{homoop}
H\left[g^{(0)}\left(u^{(0)}_i, b^{(0)}\right)\right] g^{(n)}(x^\mu,r ) = S_n \, .
\end{equation}
Here $H$ is a linear differential operator of second order
in the variable $r$ alone and contains no boundary derivatives. As $g^{(n)}$ is already of order $\epsilon^n$,
and since every boundary derivative appears with an additional power of
$\epsilon$, $H$ is an \textit{ultralocal} operator (no derivatives in $x^\mu$).
Hence $H$ is a differential operator only in the variable $r$ and  does not depend  on the variables $x^\mu$.

The precise form of this operator at a point $x^\mu_0$ depends only on the values of $u^{(0)}_i$ and $b^{(0)}$ at $x^\mu_0$ but not on the derivatives of these functions at that point. Furthermore, the operator $H$ is independent of $n$; it is the same at all orders in $\epsilon$.
The difficulty in solving Einstein's equations does not come from $H$ but from the increasing complexity of the ``source term'' $S_n$ that is an expression of $n^{{\rm th}}$ order in boundary derivatives
of $u^{(0)}_i$ and $b^{(0)}$, as well as of $(n-k)^{{\rm th}}$ order in $u_i^{(k)}$, $b^{(k)}$ for all $k \leq n-1$.

Before we proceed with our calculations we wish to comment on the general structure of the $5$-dimensional Einstein equations we are going to solve below. Einstein's equations in five dimensions correspond to $(5+1)\cdot 5/2$ independent equations of which $4$ do not involve $g^{(n)}$ and only constrain the form of expression of $b$ and $u^i$. We will call these \textit{constraint equations} and they turn out to be equivalent to
\begin{equation}
   \partial_\mu T^{\mu\nu}_{(n-1)}=0\,,
\end{equation}
where $T^{\mu\nu}_{(n-1)}$ is the boundary stress-energy tensor dual to the metric $g$ up to $\mathcal{O}(\epsilon^{n-1})$. Of the other $11$ equations, one is redundant and the 10 dynamical equations left are used to determine $g^{(n)}$ to second order in $r$.
We now make a gauge choice by imposing that the metric be of the form
\begin{equation}
   ds^2=-2u_\mu(x) S(x,r) dx^\mu dr+\chi_{\mu\nu}(x,r) dx^\mu dx^\nu\,,
\end{equation}
such that every constant $x^\mu$ trajectory corresponds to an ingoing null geodesic in $r$. The residual $SO(3)$ symmetry in the spacial boundary directions is what allows us to reduce the second order differential equations in $r$ to first order ones. The dynamical equations can then be recast into a set of first order decoupled equations that we can solve by integrating the source and thus guaranteeing regularity of the solution. The ambiguities in the integration constant can by removed by choosing the Landau gauge,
\begin{equation}
    u_\mu T^{\mu\nu}_\mathrm{dissip}=0\,,
\end{equation}
where $T^{\mu\nu}_\mathrm{dissip}$ includes all higher dissipative orders.

\subsection{Results}
Implementing this procedure to first order the metric is found to be
\begin{eqnarray}
    ds^2=-2u_\mu dx^\mu dr+r^2\left[\eta_{\mu\nu}+\left(1-f(br)\right)u_\mu u_\nu\right] dx^\mu dx^\nu \\
    + 2 r \left[ b r F(br) \sigma_{\mu\nu} + \frac{1}{3} u_\mu u_\nu \theta - \frac{1}{2} u^\rho \partial_\rho (u_\mu u_\nu) \right] dx^\mu dx^\nu\,,
\end{eqnarray}
with
\begin{eqnarray}
    F(r)\equiv \int_r^\infty \frac{x^2+x+1}{x(x+1)(x^2+1)} dx = \frac{1}{4} \left[ \ln\left(\frac{(1+r)^2(1+r^2)}{r^4}\right) - 2 \tan^{-1} r + \pi \right]\,,
\end{eqnarray}
and the stress-energy tensor
\begin{equation}
    T^{\mu\nu}=\frac{1}{b^4}\left(4u^\mu u^\nu+\eta^{\mu\nu}\right)-\frac{2}{b^3}\sigma^{\mu\nu}\,.
\end{equation}
The first line of the metric is just the zeroth order solution and the second line is the first order correction. Similarly the first term of the stress-energy tensor is the usual perfect fluid result and the second term is the dissipative correction.

The second order calculation can be carried out in the same way and the result can be found in \cite{Bhattacharyya:2008jc}. $T^{\mu\nu}_{(2)}$ has five independent (Lorentz and Weyl) covariant terms (of the form $\sigma\sigma$, $\omega\omega$, $aa$, $\dots$) from which we can read off second order fluid parameters like the relaxation time, which form a signature of conformal fluids with a gravity dual.

In \cite{Bhattacharyya:2008xc} an analysis of the event horizon of the solution yielded the entropy current on the boundary, which was shown to be never decreasing.

%
\section{Conclusions}
%

The fluid/gravity correspondence provides us with an exceptionally powerful tool for calculations and provides an interesting connection between two seemingly disconnected fields.  It is analogous to a giant ``Laplace transform'' where, when confronted with a difficult problem, we can switch to a ``space'' where the calculations can be carried out, then the results are translated back into the language of the original problem. For example, in this paper we have calculated the stress-energy tensor of a gauge theory, with dissipative corrections, by rephrasing the problem in the language of gravity, in a regime where the calculation is tractable, then translating the result back into the language of gauge theory.

In this review we have discussed a new technique that facilitates both the calculation and parameter matching. By calculating the stress-energy tensor on the boundary induced from the bulk, and then comparing that with the stress-energy of fluid dynamics, we can quickly calculate hydrodynamic properties. Because this process is iterative, dissipative corrections can be evaluated with relatively little effort.   
 
This new correspondence can be used to describe the quark-gluon plasma created in experiments of relativistic heavy ions collisions~\cite{Shuryak:2004yf,Romatschke:2007ad,Liu:2006ug,Liu:2006nn}. This has been done for the shear viscosity to entropy ratio with encouraging results, although it remains to be understood to what extent these calculations apply to real QCD and how to estimate theoretical errors~\cite{Liu:2006he}. This does not constitute experimental evidence in favour of string theory, but merely shows that string theory can give new insights into other fields of physics.

We would like to end these lecture notes with a brief discussion on the role of fluid hydrodynamics in quantum gravity. Over the last twenty years a broad class of hydrodynamic systems have been investigated, referred to as analogue models/ emergent spacetimes, whose linear excitations experience an effective metric tensor, see for example~\cite{Unruh:1981bi,Matt.-Visser:2002ot,Schutzhold:2007aa,Barcelo:2005ln}. The gravity analogy is only valid in the linear regime, beyond this its exhibits model-dependent dynamics. Any attempt to relate these toy models with Einstein gravity has only been partially successful, see for example Sakharov's induced gravity~\cite{Visser:2002hf,Barcelo:2001lu}.

The correspondence outlined exhibits a surprising relation between fluid dynamics and general relativity at a fully dynamical level. The relevance of this duality, from a conceptional viewpoint, for alternative approaches towards quantum gravity has yet to be investigated. In general we would like to close our interpretation of Veronika Hubeny's lecture series by stressing the importance of dualities such as the one studied here, as they are of interdisciplinary nature and open a window for an alternative approach towards quantum gravity.

\acknowledgments{We would like to thank Veronika Hubeny for presenting these lectures and for her comments and suggestions on these notes. Also, we wish to thank the organizers of the 2008 Summer School on Particles, Fields, and Strings held at the University of British Columbia, particularly Gordon W. Semenoff for initiating these proceedings. SW would like to thank Bill Unruh for many discussions and comments on the fluid/gravity duality.}

\bibliographystyle{JHEP}

\begin{thebibliography}{10}

\bibitem{Policastro:2001yc}
G.~Policastro, D.~T. Son, and A.~O. Starinets, {\it The shear viscosity of
  strongly coupled {N} = 4 supersymmetric {Yang}--{Mills} plasma},  {\em Phys.
  Rev. Lett.} {\bf 87} (2001) 081601,
  [\href{http://arxiv.org/abs/arXiv:hep-th/0104066}{{\tt
  arXiv:hep-th/0104066}}].

\bibitem{Bhattacharyya:2008jc}
S.~Bhattacharyya, V.~E. Hubeny, S.~Minwalla, and M.~Rangamani, {\it Nonlinear
  fluid dynamics from gravity},  {\em JHEP} {\bf 02} (2008) 045,
  [\href{http://arxiv.org/abs/arXiv:0712.2456}{{\tt arXiv:0712.2456}}].

\bibitem{Bhattacharyya:2008xc}
S.~Bhattacharyya, V.~E. Hubeny, R.~Loganayagam, G.~Mandal, S.~Minwalla,
  T.~Morita, M.~Rangamani, and H.~S. Reall, {\it Local fluid dynamical entropy
  from gravity},  {\em Journal of High Energy Physics} {\bf 2008} (2008),
  no.~06 055, [\href{http://arxiv.org/abs/arXiv:0803.2526v2}{{\tt
  arXiv:0803.2526v2}}].

\bibitem{Son:2007vk}
D.~T. Son and A.~O. Starinets, {\it Viscosity, black holes, and quantum field
  theory},  {\em Ann. Rev. Nucl. Part. Sci.} {\bf 57} (2007) 95--118,
  [\href{http://arxiv.org/abs/arXiv:0704.0240}{{\tt arXiv:0704.0240}}].

\bibitem{Aharony:1999ti}
O.~Aharony, S.~S. Gubser, J.~M. Maldacena, H.~Ooguri, and Y.~Oz, {\it Large {N}
  field theories, string theory and gravity},  {\em Phys. Rept.} {\bf 323}
  (2000) 183--386, [\href{http://arxiv.org/abs/arXiv:hep-th/9905111}{{\tt
  arXiv:hep-th/9905111}}].

\bibitem{Mateos:2007ay}
D.~Mateos, {\it String theory and quantum chromodynamics},  {\em Class. Quant.
  Grav.} {\bf 24} (2007) S713--S740,
  [\href{http://arxiv.org/abs/arXiv:0709.1523}{{\tt arXiv:0709.1523}}].

\bibitem{Horowitz:1991cd}
G.~T. Horowitz and A.~Strominger, {\it {Black strings and P-branes}},  {\em
  Nucl. Phys.} {\bf B360} (1991) 197--209.

\bibitem{Gubser:1996de}
S.~S. Gubser, I.~R. Klebanov, and A.~W. Peet, {\it Entropy and temperature of
  black 3-branes},  {\em Phys. Rev.} {\bf D54} (1996) 3915--3919,
  [\href{http://arxiv.org/abs/arXiv:hep-th/9602135}{{\tt
  arXiv:hep-th/9602135}}].

\bibitem{Hawking:1974rv}
S.~W. Hawking, {\it Black hole explosions},  {\em Nature} {\bf 248} (1974)
  30--31.

\bibitem{Hawking:1974sw}
S.~W. Hawking, {\it Particle creation by black holes},  {\em Commun. Math.
  Phys.} {\bf 43} (1975) 199--220.

\bibitem{Balasubramanian:1999re}
V.~Balasubramanian and P.~Kraus, {\it A stress tensor for anti-de {Sitter}
  gravity},  {\em Commun. Math. Phys.} {\bf 208} (1999) 413--428,
  [\href{http://arxiv.org/abs/arXiv:hep-th/9902121}{{\tt
  arXiv:hep-th/9902121}}].

\bibitem{Poisson:2004aa}
E.~Poisson, {\em A relativist's toolkit}.
\newblock Cambridge University Press, Cambridge, 2004.
\newblock The mathematics of black-hole mechanics.

\bibitem{Shuryak:2004yf}
E.~Shuryak, {\it Why does the quark gluon plasma at {RHIC} behave as a nearly
  ideal fluid?},  {\em Prog. Part. Nucl. Phys.} {\bf 53} (2004) 273--303,
  [\href{http://arxiv.org/abs/arXiv:hep-ph/0312227}{{\tt
  arXiv:hep-ph/0312227}}].

\bibitem{Romatschke:2007ad}
P.~Romatschke and U.~Romatschke, {\it Viscosity information from relativistic
  nuclear collisions: {How} perfect is the fluid observed at {RHIC}?},  {\em
  Phys. Rev. Lett.} {\bf 99} (2007) 172301,
  [\href{http://arxiv.org/abs/arXiv:0706.1522}{{\tt arXiv:0706.1522}}].

\bibitem{Liu:2006ug}
H.~Liu, K.~Rajagopal, and U.~A. Wiedemann, {\it Calculating the jet quenching
  parameter from {AdS/CFT}},  {\em Phys. Rev. Lett.} {\bf 97} (2006) 182301,
  [\href{http://arxiv.org/abs/arXiv:hep-ph/0605178}{{\tt
  arXiv:hep-ph/0605178}}].

\bibitem{Liu:2006nn}
H.~Liu, K.~Rajagopal, and U.~A. Wiedemann, {\it An {AdS/CFT} calculation of
  screening in a hot wind},  {\em Phys. Rev. Lett.} {\bf 98} (2007) 182301,
  [\href{http://arxiv.org/abs/arXiv:hep-ph/0607062}{{\tt
  arXiv:hep-ph/0607062}}].

\bibitem{Liu:2006he}
H.~Liu, K.~Rajagopal, and U.~A. Wiedemann, {\it Wilson loops in heavy ion
  collisions and their calculation in {AdS/CFT}},  {\em JHEP} {\bf 03} (2007)
  066, [\href{http://arxiv.org/abs/arXiv:hep-ph/0612168}{{\tt
  arXiv:hep-ph/0612168}}].

\bibitem{Unruh:1981bi}
W.~G. Unruh, {\it Experimental black hole evaporation},  {\em Phys. Rev. Lett.}
  {\bf 46} (1981) 1351--1353.

\bibitem{Matt.-Visser:2002ot}
M.~Novello, M.~Visser, and G.~Volovik, {\em {Artificial Black Holes}}.
\newblock World Scientific, Singapore; River Edge, U.S.A., 2002.

\bibitem{Schutzhold:2007aa}
R.~Sch{\"{u}}tzhold and W.~G. Unruh, {\it Quantum analogues: From phase
  transitions to black holes and cosmology},  {\em Lecture Notes in Physics}
  {\bf 718} (2007).

\bibitem{Barcelo:2005ln}
C.~Barcel\'o, S.~Liberati, and M.~Visser, {\it Analogue gravity},  {\em Living
  Rev. Rel.} {\bf 8} (2005) 12,
  [\href{http://arxiv.org/abs/arXiv:gr-qc/0505065}{{\tt arXiv:gr-qc/0505065}}].

\bibitem{Visser:2002hf}
M.~Visser, {\it Sakharov's induced gravity: A modern perspective},  {\em Mod.
  Phys. Lett.} {\bf A17} (2002) 977--992,
  [\href{http://arxiv.org/abs/arXiv:gr-qc/0204062}{{\tt arXiv:gr-qc/0204062}}].

\bibitem{Barcelo:2001lu}
C.~Barcel\'o, M.~Visser, and S.~Liberati, {\it Einstein gravity as an emergent
  phenomenon?},  {\em Int. J. Mod. Phys.} {\bf D10} (2001) 799--806,
  [\href{http://arxiv.org/abs/arXiv:gr-qc/0106002}{{\tt arXiv:gr-qc/0106002}}].

\end{thebibliography}
\providecommand{\href}[2]{#2}\begingroup\raggedright\endgroup

\end{document}